\documentclass[useAMS,usenatbib]{mn2e}
\usepackage{epsf}
\usepackage{amssymb}
\usepackage{graphicx}
\usepackage{eqnarray,amsmath}
\usepackage[usenames]{color}

\newcommand {\bc}{\begin {center}}
\newcommand {\ec}{\end {center}}
\newcommand {\be}{\begin {equation}}
\newcommand {\ee}{\end {equation}}
\title[Cool filaments in cluster cores]{Powering of cool filaments in
  cluster cores by buoyant bubbles. I. Qualitative model.}
\author[Churazov et al.]{E.~Churazov$^{1,2}$,  M.~Ruszkowski$^{3,4}$, A.~Schekochihin$^{5}$\\
$^{1}$ MPI f\"ur Astrophysik, Karl-Schwarzschild str. 1, Garching, D-85741, Germany\\
$^{2}$ Space Research Institute, Profsoyuznaya str. 84/32, Moscow,
  117997, Russia\\
$^3$  Department of Astronomy, The University of
Michigan, 500 Church Street, Ann Arbor, MI 48109, USA \\
$^4$ The Michigan
Center for Theoretical Physics, 3444 Randall Lab, 450 Church St, Ann
Arbor, MI 48109, USA \\
$^5$ The Rudolf Peierls Centre for Theoretical Physics, University of
  Oxford, 1 Keble Road, Oxford, OX1 3NP, UK 
 }

\begin{document}

\date{Accepted .... Received ...}

\pagerange{\pageref{firstpage}--\pageref{lastpage}} \pubyear{2012}

\maketitle

\label{firstpage}

\begin{abstract}
Cool-core clusters (e.g., Perseus or M87) often possess a network of
bright gaseous filaments, observed in radio, infrared, optical and X-ray
bands. We propose that these filaments are powered by the reconnection
of the magnetic field in the wakes of buoyant bubbles. AGN-inflated
bubbles of relativistic plasma rise buoyantly in the cluster
atmosphere, stretching and amplifying the field in the wake to values
of $\displaystyle \beta =8\pi P_{gas}/B^2\sim 1$. The
field lines in the wake have opposite directions and are forced
together as the bubble motion stretches the filament. This setup bears
strong similarity to the coronal loops on the Sun or to the Earth's
magneto-tail. The reconnection process naturally explains both the
required level of local dissipation rate in filaments and the
overall luminosity of filaments. The original source of power for the filaments
is the potential energy of buoyant bubbles, inflated by the central
AGN.
\end{abstract}
\begin{keywords} magnetic reconnection - (galaxies:) quasars: supermassive black holes - galaxies: clusters: intracluster medium - X-rays:
galaxies: clusters
\end{keywords}

\section{Introduction}
\label{sec:intro}
Networks of bright gaseous filaments are ubiquitous in the centres of
cool-core clusters \citep[e.g.,][]{2010ApJ...721.1262M}. H$_{\alpha}$ filaments around NGC1275 in the
Perseus cluster are perhaps the most famous example
\citep[e.g.,][]{1957IAUS....4..107M,1970ApJ...159L.151L}. These
filaments are observed in many bands/lines, including CO
\citep[e.g.,][]{1989ApJ...336L..13L,2006A&A...454..437S},
near-infrared (NIR) lines
\citep{2012MNRAS.426.2957M}, optical lines
\citep[e.g.,][]{2001AJ....122.2281C} and soft X-rays
\citep{2003MNRAS.344L..48F}, suggesting a multi-temperature gas sharing
approximately the same space within the cluster. For a recent summary of
observational results on NGC1275 and M87 filaments, see,
e.g., \citet{2011MNRAS.417..172F,2013ApJ...767..153W} and references
therein. Below we discuss NGC1275 and M87 collectively, under the implicit
assumption that the same universal mechanism is responsible for the filamentary
structures in both objects (and also in other cool-core clusters).

The source of energy powering the filaments is a long-standing
problem. The bolometric luminosity of the filaments in the NIR-optical band could
be at the level of 10-20\% of the total X-ray luminosity of the
cluster core. Various scenarios have been considered, including shocks
\citep{1988ApJ...329...66D}, photoionization by optical/ultraviolet or X-ray
radiation \citep[e.g.,][]{1989ApJ...338...48H,1994ApJS...95...87V},
and thermal conduction
\citep[e.g.,][]{1989MNRAS.237.1147B}. \citet{2009MNRAS.392.1475F}
argued that the spectra of the outer filaments require the line
excitation by energetic particles, although not all line ratios are
consistent with this scenario \citep{2012MNRAS.426.2957M}. Recently
\citet{2011MNRAS.417..172F} and \citet{2013ApJ...767..153W} suggested that the
filaments are powered by the hot intracluster medium (ICM), which penetrates into the
filaments via turbulent reconnection \citep[see also][]{2004A&A...422..445S}.

The filaments are long and thin, probably consisting of many threads
\citep{2007ApJ...665.1057F,2008Natur.454..968F}. This suggests that
the magnetic field is playing a role. The role of magnetic fields and
in particular magnetic reconnection as a source of energy for filaments has
been considered in, e.g., \citet{1990ApJ...348...73S,1996MNRAS.280..438J,1998AJ....116...37G}. It was assumed that an inflow of cooling gas
(in the frame of the original cooling flow model, see, e.g. \citealt{1994ARA&A..32..277F}) increases the magnetic energy density in the core of the cluster. The relative
contribution of the magnetic field to the energy density is further
amplified by the radiative cooling losses of the gas thermal energy. 

\begin{figure*}
\begin{minipage}{0.49\textwidth}
\includegraphics[trim= 0mm 0cm 0mm 0cm, width=1\textwidth,clip=t,angle=0.,scale=0.6]{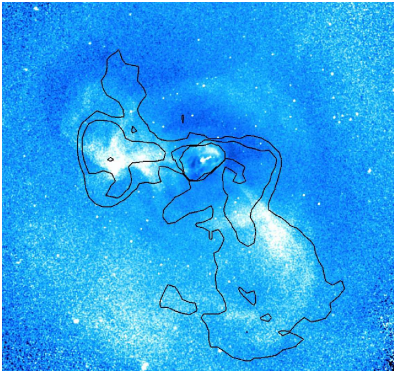}
\end{minipage}
\begin{minipage}{0.49\textwidth}
\includegraphics[trim= 1cm 6cm 0mm 9cm,width=1\textwidth,clip=t,angle=0.,scale=0.9]{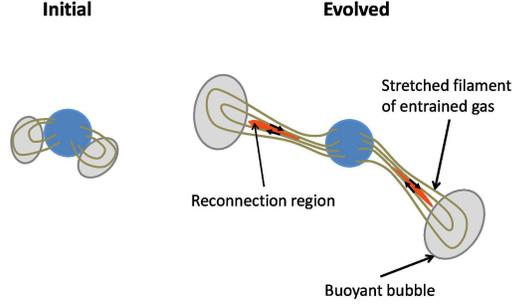}
\end{minipage}
\caption{Qualitative picture of the current sheets formation by
  AGN-inflated buoyant bubbles of relativistic plasma, rising in the
  cluster atmosphere. Nearby elliptical galaxy M87/Virgo is used in
  this example. {\bf Left:} Morphology of soft X-ray filaments in M87
  \citep{2007ApJ...665.1057F} and overall morphology of the radio
  emitting plasma \citep{2000ApJ...543..611O}, superposed as
  contours. Optical filaments are largely co-spatial with X-ray
  filaments. Buoyant bubbles rise in the atmosphere, entraining the
  low-entropy gas from the core
  \citep{2000AA...356..788C,2001ApJ...554..261C} and
  stretching/squeezing the fluid elements in the wake. Radio emission
  traces the distribution of the relativistic plasma produced by the
  AGN. {\bf Right:} Schematic evolution of the magnetic field in the
  wake. As the bubbles rise, they stretch the magnetic field lines in
  the entrained fluid elements, thus increasing the strength of the
  field. The field lines, anchored to the gas in the cluster core,
  have opposite directions in the wake. They are forced together as
  the bubble rises. This setup bears strong similarity to the coronal
  loops on the Sun or to the Earth's magneto-tail, where reconnection is
  believed (and, in some cases, observed) to take place.
\label{fig:m}
}
\end{figure*}

Many models mentioned above appeal to thermal or gravitational
  energy of the surrounding ICM as the source of energy powering the filaments.
Here we consider a different scenario, in which buoyant bubbles of
relativistic plasma stretch the magnetic field lines and drive the
fields of opposite direction together. In this model the
active galactic nucleus (AGN)-inflated bubbles provide the energy that powers the
filaments. A schematic picture of this process is shown in
Fig.~\ref{fig:m}. 

The structure of this paper is as follows. In \S\ref{sec:bub} we
briefly summarize the relevant properties of AGN-inflated bubbles. In
\S\ref{sec:rise} we discuss the amplification of the magnetic field by
the rising bubbles. In \S\ref{sec:rec} we provide an order-of-magnitude
< estimate of the rate of energy dissipation by reconnecting magnetic
< fields in the bubble's tail and the resulting luminosity of the filaments. 
 In \S\ref{sec:dis} we discuss the
overall energetics of filaments and other basic properties of our
model. Our findings are summarized in \S\ref{sec:con}.

\section{Buoyant bubbles}
\label{sec:bub}
Observations suggest that AGN activity regulates the thermal state of
the gas by injecting energy into the intra-cluster medium in the cores
of relaxed clusters, where radiative cooling time is often as short as
few times $10^8$ yr. AGN jets drive a shock into the ICM and inflate
  a cocoon of shock-heated material around the nucleus. As the size of
  the cocoon increases, the expansion velocity becomes subsonic \cite[see,
    e.g.,][]{1998ApJ...501..126H}. The cocoon transforms into one of
  several bubbles
  of relativistic plasma, whose evolution is dominated by the buoyancy
  force. The cocoon-to-bubble transformation is accompanied by the
  entrainment of lumps of ambient ICM and subsequent advection of
  these lumps. The bubbles rise buoyantly through the gaseous
atmosphere, leading to a number of spectacular phenomena such as
expanding shocks, X-ray-dim and radio-bright cavities, X-ray-dim and
radio-dim ``ghost'' cavities (aged versions of ``normal'' cavities)
and filaments of cool gas in the wakes of the rising bubbles formed by
the entrained low-entropy material from the core
\citep{2000AA...356..788C,2001ApJ...554..261C}. With Chandra and
XMM-Newton, these features are now studied in great detail in many
systems.

Observations further suggest that a large fraction of the energy output of
the AGN goes into the enthalpy of the bubbles 
$\displaystyle
H=\gamma_bPV_b/(\gamma_b-1)$, rather than into shocks. Here
$\gamma_b$ is the adiabatic index of the gas inside the bubble
($\gamma_b=4/3$ or $5/3$ depending on whether a relativistic or
a non-relativistic gas is considered), $P$ is the pressure inside the
bubble and $V_b$ is the bubble volume. The partitioning of the AGN energy
between shocks and bubble enthalpy depends on the energy
injection rate, duration of the AGN outburst and initial conditions
\citep[e.g.,][, Forman et al. in preparation]{2007ApJ...665.1057F}, but fiducial models predict that
about 70\% (and certainly more than 50\%) goes into the enthalpy of
bubbles. Furthermore, the lack of very strong shocks around observed
bubbles suggests that the thermal gas pressure of the ICM supporting
the bubble can be used in the above expression for the enthalpy $H$
(i.e., $P\approx P_{gas}$, the bubble and the ICM are in pressure balance).

The bubbles then serve as a reservoir of potential energy $\sim H$,
deposited by the AGN. The dynamics of the bubble rise is set by the
competition of the buoyancy force and the drag from the ambient
gas. Even if we consider only the hydrodynamic drag (i.e., ignoring a
possible contribution of magnetic fields) the rise velocity is
expected to be subsonic \citep{2000AA...356..788C}.  Indeed, the
buoyancy force is $F_{b}\sim\rho V_b g$, where $g$ is the gravitational
acceleration, and it is balanced by the ram pressure of the ICM
(inertial drag force)
$\displaystyle F_{ram}\sim A\rho v_{b}^2$, where $v_{b}$ is the
bubble's terminal velocity and $A$ is the cross-section of the bubble. Equating $F_{ram}$ and
$F_{b}$ gives  $\displaystyle
v_{b}\sim\sqrt{gR}$, where $R$ is the bubble radius. This
terminal velocity will be subsonic/transsonic as long as the bubble radius does
not exceed the pressure scale height of the atmosphere.  Assuming that
the bubble is moving subsonically and does not mix with the ambient
ICM, the volume of the bubble expands adiabatically $\displaystyle
V_b=V_{b,0} \left ( \frac{P}{P_0}\right ) ^{-\frac{1}{\gamma_b}}$,
where $P=P(r)$, $P_0=P(r_0)$ is the ICM pressure and $r_0$ and $r$ are
the initial and current distances of the bubble from the cluster centre.
 For simplicity, we assume below a power-law
dependence of the pressure on the radius: $\displaystyle P=P_0\left (
\frac{r}{r_0} \right )^{-\alpha}$. Typically $\alpha \sim 0.7-1$ for
the relevant range of radii in cool-core clusters. For example, using radial
density and temperature profiles from \citet{2003ApJ...590..225C} and \citet{2007ApJ...665.1057F}, we obtained $\alpha=0.8$
and 0.9 for the Perseus cluster and M87, respectively.

The ambient material and the bubble itself can be threaded by the
magnetic fields. As the bubble rises, the magnetic field is amplified, a process discussed in the next section.

\section{Rise of the bubble}
\label{sec:rise}

The role of magnetic fields in the evolution of buoyant bubbles has
been considered in, e.g., \citet{2007MNRAS.378..662R,2008MNRAS.383.1359R,2009ApJ...694.1317O}. Here
we concentrate specifically on the threads of the magnetic field in
the wake of a rising bubble.

A sketch of the configuration is shown in Fig.\ref{fig:m}. We assume
that the bubble advects a lump of the ICM threaded by magnetic field
lines, which are anchored to the gas in the cluster core. We also assume
that initially the reconnection of the magnetic field can be neglected
(this will be checked in \S\ref{sec:rec}).
As the
bubbles rise, the advected fluid elements and the magnetic field
frozen into them are
stretched by the bubble motion. We start by considering the evolution
of such an advected fluid element occupying a volume $V \lesssim V_b$. As the
fluid element moves from $r_0$ to $r$, its volume expands adiabatically
$V=V_0 \left ( \frac{P}{P_0}\right ) ^{-\frac{1}{\gamma}}$, where
$\gamma$ is the adiabatic index of the ICM.  The linear size
of the stretched fluid element along the direction of motion can be
estimated as $l\approx R_0+r-r_0$, where $R_0$ is the initial size of
the fluid element. In the limit $r \gg r_0$,
$l\sim r$. The cross-section of the fluid element in the
perpendicular direction is then
\begin{eqnarray}
A\sim\frac{V}{l}\sim\frac{V_0}{l} \left (
\frac{P}{P_0}\right ) ^{-\frac{1}{\gamma}}\sim \frac{V_0}{r} \left (
  \frac{r}{r_0} \right
)^{\frac{\alpha}{\gamma}}\propto r^{\frac{\alpha}{\gamma}-1}.
\label{eq:a}
\end{eqnarray}
Thus, for $\alpha < \gamma$, the cross-section
of the fluid element shrinks as the bubble rises. 

Here we neglect several effects that might influence the
  behaviour of rising bubbles. For instance, gas viscosity and/or
  magnetic fields may affect
  the development of instabilities at the bubble/ICM interface
  \citep[e.g.,][]{2005MNRAS.357..242R,2009ApJ...704.1309D}. We also neglect possible
  turbulence in the ICM surrounding the entrained fluid elements
  considered above. Namely, we assume that the
  stretching velocity of the fluid elements is large enough to dominate over
  the turbulent ICM motions and/or the overdensity of entrained fluid
  elements and that
  their internal magnetic fields help to prevent their
  disruption. Thus, we assume that entrained fluid elements are evolving
  solely under the action of stretching due to the bubbles and are approximately
  adiabatic. 

Other potentially important effects are associated with the magnetothermal
 \citep[MTI;][]{2000ApJ...534..420B} and heat-flux-driven buoyancy
 \citep[HBI;][]{2008ApJ...673..758Q} instabilities, operating in a
stratified weakly magnetized medium. Inside the entrained fluid elements, the magnetic
field is mostly radial, while the temperature is decreasing with
radius due to adiabatic expansion of the fluid
elements. Even if the temperature increases with radius (due to
radiative cooling), the stretching velocity is expected to be a large
fraction of the sound speed and the velocity of this magnitude is
likely to overwhelm the effects of the HBI
\citep{2010ApJ...713.1332R,2010ApJ...712L.194P}. Therefore,
these conditions are not favorable for either MTI or HBI,
which can
therefore be neglected in this qualitative study.

The stretching of the fluid element will align and amplify the
magnetic field $B$. From the conservation of the magnetic flux through
the cross-section of the advected fluid element, and using Eq.(\ref{eq:a}), we find
\begin{eqnarray}
\frac{B}{B_0}\sim\frac{A_0}{A}\sim\frac{l}{R_0}\left (\frac{P}{P_0}\right )
^{\frac{1}{\gamma}}\sim \frac{r}{R_0}\left (\frac{r}{r_0}\right )
^{-\frac{\alpha}{\gamma}},
\end{eqnarray}
where $B_0$ is the initial magnetic field and $A_0\sim V_0/R_0$ the
initial cross-section. The corresponding magnetic
energy is
\begin{eqnarray}
E_B=\frac{B^2}{8\pi}V\sim\frac{B_0^2}{8\pi}V_0\frac{l^2}{R_0^2}\left
(\frac{P}{P_0}\right ) ^{\frac{1}{\gamma}}\sim
\frac{P_0V_0}{\beta_0}\frac{r^2}{R_0^2} \left(\frac{r}{r_0}\right )
^{-\frac{\alpha}{\gamma}},
\label{eq:eb}
\end{eqnarray}
where $\displaystyle \beta_0=8\pi\frac{P_0}{B_0^2} \sim 100$
\citep[e.g.,][]{2002ARA&A..40..319C} is the $\beta$ parameter of the
ICM near the initial position of the bubble.
Using this expression, we can estimate the maximum distance $r_{max}$
from the cluster centre that the
bubble can reach -- this is the radius where the buoyancy force
$\displaystyle F_{b}\sim \rho
V_b g \sim V_b\frac{dP}{dr}\sim V_b\frac{P}{r}$ is equal to $\displaystyle F_B\sim
\frac{dE_B}{dr} \sim \frac{E_B}{r}$. In the limit of $r_{max}\gg r_0,R_0$, the equality
$F_{b}\sim F_{B}$ is reached at
\begin{eqnarray}
r_{max}\sim r_0 \left (\beta_0
\frac{V_{b,0}}{V_0} \frac{R_0^2}{r_0^2}
\right )^\frac{1}{2+\alpha-\alpha/\gamma -\alpha/\gamma_b}.
\end{eqnarray}
  At this radius, the value
of $\beta$ in the stretched fluid element~is
\begin{eqnarray}
\beta (r_{max})\sim \frac{PV}{E_B} \sim 
\beta_0 \frac{R_0^2}{r_0^2} \left (\beta_0
\frac{V_{b,0}}{V_0} \frac{R_0^2}{r_0^2}
\right )^{-\frac{2\gamma+\alpha(\gamma-2)}{2\gamma+\alpha(\gamma-1-\gamma/\gamma_b)}}.
\end{eqnarray}
Setting $V_{b,0}\sim r_0^3$ (i.e., initial bubble size is comparable
with the initial distance from the cluster centre) and the initial
volume of the fluid element $V_{0}\sim
R_0^3$, we get
\begin{eqnarray}
r_{max} \sim r_0 \left (\beta_0
  \frac{r_0}{R_0} \right
  )^\frac{1}{2+\alpha-\alpha/\gamma -\alpha/\gamma_b}\sim 10~r_0
  \left (\frac{r_0}{R_0}\right )^{0.6} \label{eq:rmax}
  \\ 
\beta (r_{max}) \sim \beta_0 \left (\beta_0 \frac{r_0}{R_0} \right
  ) ^{-\frac{2\gamma+\alpha(\gamma-2)}{2\gamma+\alpha(\gamma-1-\gamma/\gamma_b)}}
\frac{R_0^2}{r_0^2}\sim \left (\frac{r_0}{R_0}
\right )^{-3},\label{eq:betamax}
\end{eqnarray}
where the last expressions have been obtained 
for a set of fiducial values $\alpha\sim 0.85,~\gamma=5/3,~\gamma_b=4/3,~\beta_0\sim100$.
Thus, taking $r_0\sim R_0$, it is reasonable to expect the bubble to rise a distance of
order $r\sim 10~r_0$ before the buoyancy and magnetic forces come into
balance. At this point, the $\beta$ parameter in the stretched fluid
elements approaches unity. Circumstantial evidence for the magnetic field energy density comparable to the 
ICM
thermal pressure was indeed presented (based on a different argument) in
\citet{2011MNRAS.417..172F,2013ApJ...767..153W}.

\section{Reconnection in the filaments}
\label{sec:rec}

Once the
bubble is at $r_{max}$, further stretching of the field lines is not
possible. The bubble (or fluid elements attached to it) would
``hang'' on the magnetic field lines. However, the field lines in the
stretched fluid elements will have opposite directions and are forced
together by the shrinking cross-section of the filament. The configuration
bears strong similarity to the solar coronal loops \citep[e.g.,][]{1976SoPh...50...85K} or the Earth
magneto-tail \citep[e.g.,][]{2000SSRv...91..507N}, making the filament prone to
reconnection. As the 
anti-parallel field lines come together, current sheets are formed,
with an inflow of magnetic energy, which is eventually dissipated
there. The release of magnetic energy allows the bubble to rise
further.

Magnetic reconnection in both collisional (MHD) and collisionless
plasmas proceeds at a rate that is independent of Ohmic resistivity
\citep{2010PhRvL.105w5002U} or
other aspects of plasma microphysics \citep{2001PhRvL..87s5004R}.
 Namely, 
one can write a rough estimate of the magnetic energy inflow per unit
surface of the reconnecting layer as follows:
\begin{eqnarray}
L_{rec}\approx \epsilon v_A \frac{B^2}{8\pi},
\end{eqnarray}
where $\displaystyle v_A=\sqrt{\frac{B^2}{4\pi n \mu m_p}}$ is the
Alfv\'en speed, $n$ is the gas particle density, $\mu$ mean particle
atomic weight and $\epsilon$ is the dimensionless reconnection rate
between $\epsilon\sim 0.01$ for collisional plasmas
\citep{2010PhRvL.105w5002U,2012PhPl...19d2303L,2009PhPl...16k2102B,2009PhRvL.103f5004D,2009MNRAS.399L.146L}
and  $\epsilon\sim 0.1$ for collisionless ones
\citep{2001JGR...106.3715B}. 

 Using this expression, we can compare the increase of the magnetic
  energy due to stretching $\displaystyle \frac{dE_B}{dt}\sim
  F_B v_{b}$ and the
  release of the magnetic  energy due to reconnection $\displaystyle
  \left (\frac{dE_B}{dt}\right )_{rec}\sim SL_{rec}$. Here $F_B\sim
  E_B/r$, $v_b\sim \sqrt{gR}\sim c_s\sqrt{V_b^{1/3}/r}$ is the
  velocity of the bubble, $c_s$ is the speed of sound (note $v_A\sim
  c_s/\sqrt{\beta}$) and  $S$ is the
  lateral area of the filament. This area can be estimated as the
  product of the length of the filament $\sim r$ and its transverse
  size $\displaystyle \sim \sqrt{V/r}$. Using the results of
  Sec.\ref{sec:rise} and our 
  fiducial values of $\alpha,~\gamma,~\gamma_b$,  we can
  estimate the radius $r_{eq}$ where the rates of generation and dissipation of
  magnetic energy are approximately equal:
\begin{eqnarray}
r_{eq}\sim r_0 \left ( \frac{\epsilon}{\eta \sqrt{\beta_0}} \right
)^{-0.4} \left ( \frac{r_0}{R_0}\right )^{-1}\sim (6-16) r_0,
\end{eqnarray}
depending on the reconnection rate $\epsilon$.
This is close to the radius $r_{max}$
where $\beta\sim 1$. For $r<r_{eq}$, reconnection is slow relative to
field stretching, so it was reasonable in \S\ref{sec:rise} to ignore
the former.

Assuming that $\beta\sim 1$, we
can replace the magnetic energy density with the thermal energy
density $\displaystyle \frac{B^2}{8\pi}\sim nkT$ and the Alfven
velocity $v_A$ with the sound speed $\displaystyle c_s=\sqrt{\gamma
  \frac{P_{gas}}{\mu m_p}}$. This gives an order-of-magnitude estimate
of the surface influx of energy\footnote{Note that while the
   dissipation rate of the magnetic energy in the reconnection process
   does not have to be the same as the reconnection rate, it is
   reasonable to expect that they are comparable (there is some
   numerical evidence in support of this, e.g.,
   \citealt{2012PhPl...19d2303L}). In our simple estimates we have
  absorbed both the reconnection and the dissipation rate into the $\epsilon$ parameter.}:
\begin{eqnarray}
L_{rec}\approx \epsilon c_s nkT.
\end{eqnarray}
For the NGC1275 and M87 the estimates of the total emitted surface
flux by the filaments are available
\citep{2011MNRAS.417..172F,2013ApJ...767..153W}: $L_{em}\sim
10^{-2}~{\rm ergs~s^{-1}~cm^{-2}}\sim 0.2 c_s nkT$ and $\sim 2.2~10^{-3}~{\rm
  ergs~s^{-1}~cm^{-2}}\sim 0.1 c_s nkT$ respectively.   Thus
there is an interesting order-of-magnitude agreement between the amount of
energy that can be produced by fast reconnection and the amount of
energy emitted by the filaments, i.e.  $\displaystyle
\frac{L_{rec}}{L_{em}}\approx \frac{\epsilon}{0.1}$. If the reconnection
rate is on the stronger side of the possible values, viz. $\epsilon\sim
0.1$, the energy release from reconnection is comparable to the cooling
losses of the filaments. 

 The evolution of buoyant bubbles in a magnetized ICM has  been
  considered in several numerical simulations
  \citep[e.g.,][]{2004ApJ...601..621R,2007MNRAS.378..662R,2008MNRAS.383.1359R,2009ApJ...694.1317O,2009ApJ...704.1309D}. Most
  of these studies are focused on the overall dynamics of the rising
  bubbles, rather than on the structure of the magnetic field in the
  wake and associated reconnection. We will address these issues in
  subsequent publications. Nevertheless, some of the features relevant
  for our discussion can be found in existing simulations, especially
  in configurations where magnetic field lines are threading the
  bubble and are anchored to the ambient ICM. For example, enhanced fields in the
  wake are seen in the 2D simulations of \citet{2004ApJ...601..621R}
  with an initially horizontal magnetic field. In
  \citet{2007MNRAS.378..662R}, the wake behind the bubble shows a narrow
  layer of close-to-zero magnetic field which is likely the area
  of anti-parallel magnetic fields, where reconnection is likely to
  happen. 

\section{Discussion}
\label{sec:dis}
The overall energetics of the cool cores are believed to be determined by
the balance of the AGN activity and gas cooling. In other words, one can assume
that the cooling losses are approximately matched by the amount of
mechanical energy pumped by the AGN into the gas in the form of
relativistic bubbles. Observations suggest that a significant (if not
dominant) fraction
 of the AGN energy goes into the enthalpy of the
bubbles rather than into shocks \citep[e.g.,][]{2002MNRAS.332..729C}. This means that potential energy of
underdense bubbles created by the AGN per unit time approximately matches 
gas cooling losses. The estimates in \S\ref{sec:rise} suggest that
by the time  $\beta$ reaches 1, the bubble has moved to
$r_{max}\sim 10 r_0$. Let us
estimate the ratio $f_B$ of the magnetic energy of the
stretched fluid element at this moment to the initial enthalpy of the bubble $H_0=\frac{\gamma_b}{\gamma_b-1}P_0V_{b,0}$:
\begin{eqnarray}
f_B=\frac{E_B(r_{max})}{H_0} \sim \frac{\gamma_b-1}{\gamma_b}
\frac{1}{\beta(r_{max})}\frac{PV}{P_0V_{b,0}} &\sim& \nonumber \\ \frac{\gamma_b-1}{\gamma_b}
\frac{1}{\beta(r_{max})} \left (\frac{r_{max}}{r_0}\right
)^{-\alpha\frac{\gamma_b-1}{\gamma_b}} \left
(\frac{R_0}{r_0}\right)^{3} &\sim& 0.1,
\end{eqnarray}
using eq.(\ref{eq:eb},\ref{eq:betamax},\ref{eq:rmax}) and neglecting
 dependence on $R_0/r_0$.
At the same time, the fraction of  enthalpy remaining in
 the bubble is
\begin{eqnarray}
f_H=\frac{H(r_{max})}{H_0}=\frac{PV_b}{P_0V_{b,0}} \sim \left (\frac{r_{max}}{r_0}\right )
^{-\alpha\frac{\gamma_b-1}{\gamma_b}}\sim 0.5.
\end{eqnarray}
The rest of the initial enthalpy has already been transferred to the
gas via hydrodynamic drag, potential energy of the uplifted gas,
magnetic energy, and excitation of $g$-modes, which then dissipate in
the ICM \citep{2002MNRAS.332..729C}. Comparison of $f_H$ and $f_B$
suggests that by the time the bubble
reaches $r_{max}$, about 20\% of its available energy will have gone
into magnetic energy forced into its tail.
The luminosity of the filaments from NIR to optical bands amounts to
10-20\% of the bolometric luminosity of the cluster cores. This means
that 10-20\% of the potential energy available conversion into
magnetic energy should indeed go into reconnection and the associated heating. When the reconnection
releases magnetic energy, the bubble continues to rise beyond
$r_{max}$. 

Note that it very likely that in real clusters, there is a considerable
spread in the values of initial parameters, such as, e.g., $\beta_0$,
$r_0/R_0$. This suggests a large variation in the appearance of the
filaments in different clusters or even of filaments/bubbles in the
same cluster.

In our simple scenario, only the regions where the configuration of
magnetic field is favorable to reconnection are observed as
filaments. The energy of the magnetic field is the principal source of
energy for the outer filaments, rather than ICM thermal energy or
photoionization, although both can contribute. The implication is that
the filaments need not necessarily grow in mass with time, instead they
thermalize and emit the bubble energy mediated by magnetic fields.

The cooling of the gas is not the central element of our model
\citep[cf.][]{1990ApJ...348...73S,1996MNRAS.280..438J} in the sense
that the main driver of the reconnection is the stretching of the field lines
by the bubbles rather then the loss of thermal energy by the cooling
gas. It is nevertheless clear that a large amount of cool gas is present
in these systems. In the simplest scenario, this gas intercepts,
thermalizes and re-emits the released energy of the magnetic
field. The discussion of how the magnetic energy is split between
the kinetic energy of the gas, its thermal energy and non-thermal particles,
or of the actual excitation of the optical lines is beyond the scope of
this Letter. We nevertheless note that the presence of non-thermal
particles may help explain many properties of the emission spectra
\citep[][]{2009MNRAS.392.1475F}.

As a speculative extension of our qualitative model, we note that the
gas leaving the reconnection region will have velocities of order
$v_A$, i.e., close to the sound speed (because $\beta\sim 1$). The outflow is
typically bi-directional, i.e., along the filaments. Reconnection in
extended current sheets is typically accompanied by generation and
ejection of copious numbers of plasmoids
\citep{2012PhPl...19d2303L,2009PhPl...16k2102B,2009PhRvL.103f5004D,2009PhRvL.103j5004S},
some of them very large  \citep{2012PhPl...19d2303L}. If the filament is
aligned along the line of sight towards an observer, this may lead to
the appearance of gas lumps moving towards the core, away from the
observer with the speed that can be as large as $\sim 10^3~{\rm km~s^{-1}}$ for the hot
gas. There is a so-called High Velocity system (HV) in the core of the
Perseus cluster -- a line-emitting region in the core of NGC1275, with
a recession velocity $\sim$3000 km/s larger than the systemic
velocity of NGC1275 \citep[e.g.,][]{1957IAUS....4..107M}, which is
nevertheless located in front of NGC1275
\citep[e.g.,][]{1973ApJ...185..809D} and, therefore, is moving towards
the nucleus. While the infall velocity of HV seems to be too large
for a conceivable plasmoid-ejection mechanism, it is nevertheless interesting to note that in
some favorable configurations, high-velocity gas lumps can be observed.

\section{Conclusions}
\label{sec:con}
We argue that buoyant bubbles in the cores of galaxy clusters stretch
the fluid elements advected from the core, forming gaseous filaments
and aligning and amplifying the magnetic field in these filaments. The
field grows to $\beta\sim 1$ after the bubbles rise a distance of the
order of 
10 times their initial size. The field lines in the wake of the bubble are anti-parallel and
are forced together.  This setup bears strong similarity to the coronal
loops on the Sun or to the Earth's magneto-tail. The reconnection process
can naturally explain both the required  local dissipation
rate in filaments  and the overall energy balance. In this model, the
original source of power for the filaments is the potential energy of
buoyant bubbles, inflated by the central AGN. Of the order of 10\% of the
total mechanical energy deposited by the AGN in the form of
such relativistic bubbles can be converted into the emission from the
filaments.

\section{Acknowledgments} 
EC acknowledges useful discussions with A.Petrukovich and H.Spruit. 
This work was supported in part by the Leverhulme Trust Network on Magnetized 
Plasma Turbulence and the programme OFN-17 of the Division of Physical
Sciences of the Russian Academy of Science. 
MR acknowledges NSF grant AST 1008454 and NASA ATP grant 12-ATP12-0017.

\label{lastpage}
\end{document}